# *Optical instabilities and spontaneous light emission by polarizable moving matter*

*Mário G. Silveirinha*[*]

[(1)]University of Coimbra, Department of Electrical Engineering – Instituto de Telecomunicações, Portugal, mario.silveirinha@co.it.pt

**Abstract**

One of the most extraordinary manifestations of the coupling of the electromagnetic field and matter is the emission of light by *charged* particles passing through a dielectric medium: the Vavilov-Cherenkov effect. Here, it is theoretically predicted that a related phenomenon may be observed when *neutral* fast polarizable particles travel near a metal surface supporting surface plasmon polaritons. Based on a *classical* formalism, it is found that at some critical velocity, even if the initial optical field is vanishingly small, the system may become unstable and may start spontaneously emitting light such that in some initial time window the electromagnetic field grows exponentially with time.

---

[*] To whom correspondence should be addressed: E-mail: mario.silveirinha@co.it.pt

# I. Introduction

The interactions between light and matter are observed in many forms: the emission, absorption, and scattering of light, optical forces on nanoparticles, Raman scattering, just to name a few. In particular, the generation of light by charged particles passing either through a medium or near by a diffraction grating has been demonstrated by Vavilov & Cherenkov [1, 2] and by Smith & Purcell [3], respectively, and has important applications in the detection of high-energy charged particles in astrophysics and particle physics. These remarkable phenomena can be explained in the framework of classical electrodynamics because a modulated beam of moving charged particles corresponds to a time varying current, leading thus to the emission of light.

On the other hand, the emission of light due to fast changes in the geometry of electrically neutral macroscopic bodies (e.g. due to the accelerated motion of material boundaries) has also been extensively discussed in quantum physics in the context of the dynamical Casimir effect [4]-[9]. Moreover, several authors predicted the emission of radiation either by bodies in relative translational motion or by rotating objects [10]-[23]. These effects are understood as being *intrinsic* to quantum electrodynamics, and no classical analogue has been reported.

The playground for this work is the scenario wherein a neutral particle – *with no net electric charge*, e.g. an electric dipole – moves closely parallel to an uncharged metal surface. According to classical theory, the dipole motion should be totally unaffected by the presence of the metal surface, provided the electromagnetic field vanishes and the dipole is in its "ground state" (classical electric dipole moment is zero) so that there are no charge oscillations. Indeed, a beam of neutral particles is supposedly uncoupled from the radiation field. Here, it is shown that, extraordinarily, if the velocity of the electric dipole is sufficiently large, a system instability may be developed so that the dipole oscillations and light emission can be triggered by vanishingly small optical noise. We note that related problems have been studied in the framework of quantum theory, in connection with the problem of quantum friction [10]-[23]. In particular, related optical instabilities have been recently linked to quantum friction in the case of nondispersive dielectric slabs [24-26] (see also Ref. [23]). Quite differently, here our analysis is fully *classical*, and both the electromagnetic field and the dynamics of the pertinent moving bodies are treated with classical theories.

## II. Natural oscillations of an electric dipole above a moving half-space

Let us consider a vertical electric dipole (oriented along the *z*-direction) standing in free-space at a distance *d* from a planar thick metallic region [Fig. 1a]. The relative velocity between the dipole and the metallic region is *v*, and is assumed time independent except if stated differently. As discussed later, having *v* independent of time may require some external action to counterbalance optically induced forces. For convenience, we take the dipole rest frame as the reference frame wherein all the physical quantities are defined. To keep the formalism simple, it is supposed that the dipole can only vibrate along the *z*-direction (anisotropic particle), but as discussed ahead the theory can be readily extended to the general case where the dipole response is isotropic.

First, we characterize the fields radiated by the dipole when it oscillates with frequency $\omega$. The electromagnetic fields in the $z > 0$ region [Fig. 1a] are the superposition of the field radiated by the dipole in free-space and the field scattered by the moving metallic slab, $\mathbf{E} = \mathbf{E}^{inc} + \mathbf{E}^s$. The "incident" electric field is given by $\mathbf{E}^{inc} = \nabla \times \nabla \times \left( \dfrac{\mathbf{p}_e}{\varepsilon_0} \Phi_0 \right)$ for $\mathbf{r} \neq \mathbf{r}'$, where $\mathbf{r}' = (0,0,d)$ is the position of the electric dipole, $\mathbf{p}_e = p_e \hat{\mathbf{z}}$ is the electric dipole moment,

$$\Phi_0 = \frac{e^{ik_0|\mathbf{r}-\mathbf{r}'|}}{4\pi|\mathbf{r}-\mathbf{r}'|} = \frac{1}{(2\pi)^2} \iint dk_x dk_y \frac{e^{-\gamma_0|z-d|}}{2\gamma_0} e^{i(k_x x + k_y y)} \tag{1}$$

is the Hertz potential, $k_0 = \omega/c$ is the free-space wave number, and $\gamma_0 = \sqrt{k_x^2 + k_y^2 - \omega^2 \varepsilon_0 \mu_0}$. In particular, the transverse incident electric field, $\mathbf{E}_t^{inc} = E_x^{inc} \hat{\mathbf{x}} + E_y^{inc} \hat{\mathbf{y}} = \nabla_t \dfrac{\partial}{\partial z} \Phi_0 \dfrac{p_e}{\varepsilon_0}$, at the interface with the moving medium can be written as a plane wave superposition as

$$\mathbf{E}_t^{inc}\Big|_{z=0} = \frac{p_e}{\varepsilon_0} \frac{1}{(2\pi)^2} \iint dk_x dk_y \frac{i\mathbf{k}_t}{2} e^{-\gamma_0 d} e^{i(k_x x + k_y y)}, \tag{2}$$

where $\mathbf{k}_t = k_x \hat{\mathbf{x}} + k_y \hat{\mathbf{y}}$. Let $\mathbf{R}$ be a (2×2) reflection matrix such that the transverse components of the scattered wave $\mathbf{E}_t^s = E_x^s \hat{\mathbf{x}} + E_y^s \hat{\mathbf{y}}$ are related to the transverse components of an incident plane wave as $\mathbf{E}_t^s = \mathbf{R}(\omega, k_x, k_y) \cdot \mathbf{E}_t^{inc}$. Then, the transverse scattered field is given by:

$$\mathbf{E}_t^s = \frac{p_e}{\varepsilon_0} \frac{1}{(2\pi)^2} \iint dk_x dk_y \mathbf{R}(\omega, k_x, k_y) \cdot \frac{i\mathbf{k}_t}{2} e^{-\gamma_0(d+z)} e^{i(k_x x + k_y y)}, \quad z > 0. \tag{3}$$

Using $\dfrac{\partial E_z^s}{\partial z} = -\nabla_t \cdot \mathbf{E}_t^s$ it is found that the $z$-component of the scattered field is:

$$E_z^s = -\frac{p_e}{\varepsilon_0}\frac{1}{(2\pi)^2}\iint dk_x dk_y \mathbf{k}_t \cdot \mathbf{R}(\omega,k_x,k_y)\cdot \mathbf{k}_t \frac{1}{2\gamma_0} e^{-\gamma_0(d+z)} e^{i(k_x x + k_y y)}, \quad z>0. \quad (4)$$

Suppose now that the dipole is characterized by the electric polarizability $\alpha_e$ such that $p_e = \varepsilon_0 \alpha_e E_{loc,z}$, being $E_{loc,z} = E_z^s + E_z^{ext}$ the local field acting on the particle and $E_z^{ext}$ the field due to some hypothetical external optical excitation. In general, the dipole oscillations are driven by the external field, and the electric dipole moment satisfies

$$\left(\alpha_e^{-1} - C_{int}\right)\frac{p_e}{\varepsilon_0} = E_z^{ext} \quad (5)$$

where the interaction constant $C_{int}$ is such that $E_z^s(\mathbf{r}') = C_{int} p_e / \varepsilon_0$, i.e.

$$C_{int} = \frac{1}{(2\pi)^2}\iint dk_x dk_y \left[-\mathbf{k}_t \cdot \mathbf{R}(\omega,k_x,k_y)\cdot \mathbf{k}_t\right]\frac{1}{2\gamma_0} e^{-2\gamma_0 d}. \quad (6)$$

Here we are interested in the natural oscillations of the system with $E_z^{ext} = 0$. Clearly, the natural oscillations occur for frequencies $\omega$ such that:

$$\alpha_e^{-1}(\omega) - C_{int}(\omega) = 0 \quad . \quad (7)$$

Because of the radiation loss, manifested in the dynamics of the dipole oscillations by means of the Abraham-Lorentz self-force [27], the electric polarizability of a dipole oscillator is required to satisfy (in case there are no other mechanisms of loss) the Sipe-Kranendonk condition [28]

$$\text{Im}\{\alpha_e^{-1}\} = -\frac{1}{6\pi}\left(\frac{\omega}{c}\right)^3, \quad (8)$$

and thus is complex-valued. Hence, typically Eq. (7) only has solutions with natural frequencies $\omega = \omega' + i\omega''$ complex valued with $\omega'' < 0$. These solutions correspond to damped oscillations ($e^{-i\omega t} = e^{-i\omega' t} e^{\omega'' t}$) such that the energy of the dipole is converted into radiation. For example, suppose that the dipole polarizability is described by a standard Lorentz-type dispersion model

$$\alpha_e^{-1} = \frac{1}{6\pi}\left(\frac{\omega_0}{c}\right)^3 \left[Q\left(1-\frac{\omega^2}{\omega_0^2}\right) - i\left(\frac{\omega}{\omega_0}\right)^3\right], \qquad (9)$$

with a resonant response at $\omega = \omega_0$ and quality factor $Q$. Note that $\text{Im}\{\alpha_e^{-1}\}$ satisfies the Sipe-Kranendonk condition. For resonators with a large quality factor $Q \gg 1$, the solution of Eq. (7) can be found with perturbation theory:

$$\omega \approx \omega_0 \left[1 + i\frac{1}{2Q}\left(-6\pi\left(\frac{c}{\omega_0}\right)^3 \text{Im}\{C_{\text{int}}\}\Big|_{\omega=\omega_0} - 1\right)\right]. \qquad (10)$$

The sign of the term in the inner brackets is the same as the sign of $\text{Im}\{\alpha_e^{-1}(\omega_0) - C_{\text{int}}(\omega_0)\}$. In particular, if the dipole is sufficiently distant from the moving medium $C_{\text{int}}$ is negligible and $\text{Im}\{\alpha_e^{-1}(\omega_0) - C_{\text{int}}(\omega_0)\} < 0$, so that the oscillations are indeed damped.

## III. Compensation of the radiation loss

Surprisingly, it is shown next that when the relative velocity between the dipole and the moving medium is sufficiently large and is kept constant in the time window of interest, the system may become unstable and support exponentially growing oscillations ($\omega'' > 0$).

For simplicity, first we use a non-relativistic approximation ($v/c \ll 1$) to obtain $\mathbf{k}_t \cdot \mathbf{R}(\omega, k_x, k_y) \cdot \mathbf{k}_t \approx k_\parallel^2 R_{p,co}(\omega - vk_x, k_x, k_y)$, being $k_\parallel^2 = \mathbf{k}_t \cdot \mathbf{k}_t$ and $R_{p,co}$ is the reflection coefficient for *p*-polarized waves calculated in the frame co-moving with the metal slab (see Appendix A). Note that $R_{p,co}$ is evaluated at the Doppler shifted frequency $\tilde{\omega} = \omega - vk_x$ and can be written explicitly as

$$R_{p,co} = \frac{\varepsilon_0 \gamma_m - \varepsilon_m \gamma_0}{\varepsilon_0 \gamma_m + \varepsilon_m \gamma_0}, \quad \text{with} \quad \gamma_m = \sqrt{k_x^2 + k_y^2 - \omega^2 \varepsilon_m \mu_0} \,. \tag{11}$$

When the effects of time retardation are neglected, $\gamma_m \approx \gamma_0 \approx k_\parallel$, we have:

$$R_{p,co} \approx \frac{\varepsilon_0 - \varepsilon_m}{\varepsilon_0 + \varepsilon_m}. \tag{12}$$

Within this approximation, we obtain the following explicit formula for the interaction constant:

$$C_{int}(\omega) \approx \frac{1}{(2\pi)^2} \int\int dk_x dk_y \, \frac{\varepsilon_m(\tilde{\omega}) - \varepsilon_0}{\varepsilon_m(\tilde{\omega}) + \varepsilon_0} \frac{k_\parallel}{2} e^{-2k_\parallel d}. \tag{13}$$

In what follows, it is assumed that the metal permittivity has a Drude type dispersion $\varepsilon_m / \varepsilon_0 = 1 - 2\omega_{sp}^2 / [\omega(\omega + i\Gamma)]$, with $\Gamma > 0$ the collision frequency and $\omega_{sp}$ the frequency of the surface plasmon resonance such that $\text{Re}\{\varepsilon_m(\omega_{sp})/\varepsilon_0\} = -1$.

Figure 2a and 2b depict the real and imaginary parts of $C_{int}$ as a function of frequency for a point dipole placed at a distance $d = 1nm$ above a thick silver slab moving with relative velocity $v = 0.07c$. In the simulations we used $\omega_{sp}/2\pi = 646\,\text{THz}$ [29] and considered different values for $\Gamma$ to model the effect of metal absorption. The curves associated with $\Gamma = 0.2\omega_{sp}$ are expected to model realistic silver loss at $\omega \approx \omega_{sp}$ [29]. The solid lines in Fig. 2a-2b were computed using Eq. (13), whereas the dashed lines were obtained based on the exact relativistic formulation and taking into account the effects of time retardation [Eqs. (6) and (A7)]. As seen, even in presence of strong material loss the imaginary part of $\text{Im}\{C_{int}\}d^3$ can be negative in a relatively wide frequency range [see Fig. 2b]. Very importantly, negative values of $\text{Im}\{C_{int}\}d^3$ imply a radiation loss reduction and that the excited surface plasmon polaritons (SPP) return to the dipole some of the radiated power [Eq. (7)].

Remarkably, in this example the radiation loss can be over-compensated, i.e. $\text{Im}\{\alpha_e^{-1}(\omega) - C_{int}(\omega)\} > 0$ because $\text{Im}\{\alpha_e^{-1}\}$ (purple line in Fig. 2b) is typically well above the lines representing $\text{Im}\{C_{int}\}$. To understand the consequences of this intriguing property, let us suppose that the dipole response is modeled by Eq. (9). Desirably, the resonance quality factor $Q$ must be much larger than unity, and the mass of the dipole should be small so that it can be accelerated to large velocities. Without loss of generality, we consider that $\omega_0 = 0.69\omega_{sp}$ (i.e. $\lambda_0 = 671nm$) and that $Q = 7.6 \times 10^7$. These parameters may model the semi-classical response of a neutral lithium atom (Li I) [30, 31, 32]. It is important to emphasize that our Lorentz dispersion model is completely classical. Feeding these values to Eq. (10) and supposing that $d = 1nm$ and $v = 0.07c$, it is found that the dipole supports natural oscillations characterized by $\omega \approx \omega_0 \left(1 + i1.1 \times 10^{-5}\right)$ when it stands above a thick silver film ($\Gamma = 0.2\omega_{sp}$). Hence, quite dramatically, our theory indicates that when the radiation loss is over-compensated the dipole may support *free* oscillations characterized by a complex frequency $\omega = \omega' + i\omega''$ with $\omega'' > 0$, so that the system becomes unstable and the oscillations tend to grow exponentially with time!

-8-

To unveil the conditions under which it is possible to have this optical instability, we derive an explicit formula for $\text{Im}\{C_{int}\}$ in the limit of vanishing material loss ($\Gamma \to 0^+$). In this limit one has

$$\text{Im}\left\{\frac{\varepsilon_m(\tilde{\omega}) - \varepsilon_0}{\varepsilon_m(\tilde{\omega}) + \varepsilon_0}\right\} = \frac{\omega_{sp}\pi}{2}\left[\delta(\tilde{\omega} - \omega_{sp}) - \delta(\tilde{\omega} + \omega_{sp})\right]. \tag{14}$$

Thus, from Eq. (13) and using $\tilde{\omega} = \omega - vk_x$ it is easily found that in the lossless case:

$$\text{Im}\{C_{int}(\omega)\} = \frac{1}{d^3}\left[G\left(\frac{\omega d}{|v|}, \frac{\omega_{sp}d}{|v|}\right) - G\left(-\frac{\omega d}{|v|}, \frac{\omega_{sp}d}{|v|}\right)\right], \quad \Gamma = 0^+ \tag{15}$$

with

$$G(a,b) = \frac{1}{8\pi}b\int_0^\infty du\sqrt{u^2 + (a-b)^2}\, e^{-2\sqrt{u^2 + (a-b)^2}}. \tag{16}$$

Figure 2c shows $\text{Im}\{C_{int}\}d^3$ as a function of $\omega_{sp}d/|v|$ for different values of $\omega$. As seen, for sufficiently small values of $\omega_{sp}d/|v|$ the imaginary part of $\text{Im}\{C_{int}\}d^3$ can be negative. The minimum value of $\text{Im}\{C_{int}\} \approx -10^{-4}/d^3$ occurs at the frequency of oscillation $\omega = \omega_{sp}$ (green solid line), i.e. at the surface plasmon resonance, and for $\omega_{sp}d/|v| \approx 0.19$. Thus, the optimal relative velocity is $v = 5.26\omega_{sp}d$. By decreasing the distance $d$ between the dipole and the moving surface, the absolute value of $\text{Im}\{C_{int}\} < 0$ can be made arbitrarily large. For $\omega = \omega_{sp}$, $v = 5.26\omega_{sp}d$, and $\Gamma = 0^+$ the condition to have over-compensation of the radiation loss is $-1/6\pi(\omega_{sp}/c)^3 + 10^{-4}/d^3 > 0$ and this yields $\omega_{sp}d/c < 0.12$. For silver films this condition imposes that $d < 8.9nm$, but for other materials with lower plasma resonance frequencies (e.g. semiconductors or graphene) greater distances may be allowed. It is shown in Appendix B, that when the dipole response is isotropic slightly larger distances may be admissible. Thus, to over-compensate the radiation loss it is required that the dipole is sufficiently near the moving surface, and in addition that the velocity is sufficiently large (ideally $v = 5.26\omega_{sp}d$). In the limit $v \to 0$, $\text{Im}\{C_{int}\}$ is always positive. For velocities larger than the optimal value, the coupling between the dipole and the surface plasmons is less effective, and thus the dipole is required to move closer to the metal surface in order that the radiation loss can be over-compensated. Figure 2d shows $\text{Im}\{C_{int}(\omega_{sp})\}$ as function of the collision frequency $\Gamma$, further demonstrating that even in presence of strong loss it is possible to have $\text{Im}\{C_{int}\}$ negative.

## IV. Conditions for an optical instability

As discussed next, the optical instabilities can be understood as being the result of the hybridization of the dipole resonance ($\omega \approx \omega_0$) and the SPPs supported by the silver slab with *negative* frequencies ($\tilde{\omega} \approx -\omega_{sp}$) in the metal co-moving frame. Here $\tilde{\omega}$ is the Doppler shifted frequency, i.e. the frequency in the frame co-moving with the silver slab. For simplicity, in this discussion a vanishing material loss is assumed ($\Gamma = 0^+$). In the near-field approximation, the guided modes of the silver slab (i.e. the SPPs) occur for $\tilde{\omega} = \pm\omega_{sp}$ which correspond to the poles of the reflection coefficient when $\Gamma = 0^+$ [Eq. (12)]. The interaction of a high $Q$ dipole with the silver slab is completely characterized by $\text{Im}\{C_{int}(\omega_0)\}$. Clearly, from Eq. (14) $\text{Im}\{C_{int}(\omega_0)\}$ is determined by two interactions with opposite effects: (*i*) the interaction with SPPs with $\tilde{\omega} = \omega_{sp}$, $k_x = \frac{\omega_0 - \omega_{sp}}{v}$ and $k_y$ arbitrary; (*ii*) the interaction with SPPs with $\tilde{\omega} = -\omega_{sp}$, $k_x = \frac{\omega_0 + \omega_{sp}}{v}$, and $k_y$ arbitrary. Hence, in general the oscillations of the dipole are determined by the hybridization of the dipole resonance with surface plasmons with either positive $\tilde{\omega} = \omega_{sp}$ or negative frequencies $\tilde{\omega} = -\omega_{sp}$. As seen from Eq. (13), for small *d* the strength of each interaction is determined by the factor $k_\| = \sqrt{k_x^2 + k_y^2}$. Thus, the interaction with negative frequencies ($\tilde{\omega} = -\omega_{sp}$) may dominate when $\omega_0 \approx \omega_{sp}$ because surface plasmons with $\tilde{\omega} = \omega_{sp}$ and $k_x = 0$ interact weakly with the dipole. In summary, for a sufficiently small distance *d* the oscillations of the dipole are determined by the hybridization of the dipolar resonance ($\omega = \omega_0$) with surface plasmon polaritons with negative frequencies $\tilde{\omega} \approx -\omega_{sp}$. It is this interaction between positive and negative frequencies that creates the opportunity to have the system instabilities and an exponentially growing oscillation.

More generally, one could consider the interaction of a dipole with a generic moving waveguide, for example a finite thickness dielectric slab. In this case, it is possible to write in the non-relativistic regime:

$$C_{int}(\omega) = \frac{-1}{(2\pi)^2} \iint dk_x dk_y R_{p,co}(\tilde{\omega}, k_x, k_y) \frac{k_\|^2}{2\gamma_0} e^{-2\gamma_0 d}, \tag{17}$$

with $\tilde{\omega} = \omega - vk_x$. For a passive material the reflection coefficient is bound to the restriction:

$$-\omega \operatorname{Im}\{R_{p,co}(\omega, k_x, k_y)\} > 0, \quad \text{when } k_x^2 + k_y^2 - \omega^2/c^2 > 0. \tag{18}$$

To prove this result, we note that if $R_{p,co}$ is the reflection coefficient for the electric field under plane wave excitation, then for an incident field with amplitude $E^{inc}$ the total transverse electric field in the vacuum region is $E = (1 + R_{p,co}) E^{inc}$ while the transverse magnetic field is $\eta_0 H = Y^p (1 - R_{p,co}) E^{inc}$, where, $\eta_0$ is the vacuum impedance, and $Y^p = \frac{1}{i\gamma_0} \frac{\omega}{c}$ is the normalized wave admittance for $p$-polarized waves in the vacuum. In particular, the component of the time-averaged Poynting vector flowing towards the interface along the normal direction is

$$S_{av} = \frac{|E^{inc}|^2}{2\eta_0} \mathrm{Re}\left\{ Y^{p*} (1 + R_{p,co})(1 - R^*_{p,co}) \right\}. \tag{19}$$

For a passive material at rest one must impose that $S_{av} \geq 0$. Taking into account that $Y^p$ is pure imaginary when $\gamma_0 = \sqrt{k_x^2 + k_y^2 - \omega^2 \varepsilon_0 \mu_0}$ is positive, we immediately arrive at the condition (18). This proves that when $v = 0$ the interactions of the dipole with evanescent waves (waves with $\gamma_0 > 0$) give a positive (negative) contribution to $\mathrm{Im}\{C_{int}(\omega)\}$ when $\omega$ is positive (negative). In other words, when $v = 0$ the interactions of the dipole with the near-field contribute to increase the loss and to reduce the lifetime of the dipole oscillations.

Notably, for a moving system the picture can change significantly. Indeed, from Eqs. (17)-(18) the incident evanescent waves associated with a transverse wave vector such that the Doppler shifted frequency $\tilde{\omega} = \omega - v k_x$ has a sign different from $\omega$ give a negative (positive) contribution to $\mathrm{Im}\{C_{int}(\omega)\}$ when $\omega$ is positive (negative), and thus contribute to *increase* the lifetime of an oscillation.

Moreover, in the lossless limit the main contribution to $\text{Im}\{C_{int}(\omega)\}$ comes from the poles of $R_{p,co}(\tilde{\omega}, k_x, k_y)$, similar to the case of SPPs [see Eq. (13) and (14)]. It is well-known that the poles of the reflection coefficient determine the dispersion of the guided modes $\omega_g = \omega_g(k_x, k_y)$. Hence, to have $\text{Im}\{C_{int}(\omega)\} \neq 0$ it is required that $\tilde{\omega} = \omega_g(k_x, k_y)$, i.e. the condition

$$\omega = \omega_g(k_x, k_y) + vk_x. \tag{20}$$

needs to be satisfied for some guided mode branch. A certain guided mode gives a negative (positive) contribution to $\text{Im}\{C_{int}(\omega)\}$ when $\omega_g \omega < 0$ ($\omega_g \omega > 0$). Note that the modal branches can be associated with either positive or negative frequencies. Indeed, because of the reality of the fields if $\tilde{\omega} = \omega_g(k_x, k_y)$ is a guided mode branch, then $\tilde{\omega} = -\omega_g(-k_x, -k_y)$ also is.

The minimum requirement to have an optical instability is that the condition (20) is satisfied for some guided mode with $\omega = \omega_0$ and that $\omega_g \omega_0 < 0$, where $\omega_0$ is the resonant frequency of the dipole. This yields the selection rules:

$$\omega_0 = \omega_g(k_x, k_y) + vk_x, \quad \text{and} \quad \omega_g \omega_0 < 0. \tag{21}$$

Even though necessary, these conditions are *insufficient* to guarantee the emergence of an instability because the gain provided by an interaction between the dipole and a guided mode that satisfies (21) must also be sufficiently large to supplant all the loss channels (e.g. the radiation loss of the dipole, and the interactions with other guided modes that satisfy the first condition in (21) but have $\omega_g \omega_0 > 0$).

From the selection rule (21), one also sees that for an interaction between positive and negative frequencies $vk_x > 0$. Moreover, for $k_y = 0$ we see that:

$$|v| = \left|\frac{\omega_0}{k_x} - \frac{\omega_g(k_x)}{k_x}\right| = \left|\frac{\omega_0}{k_x}\right| + \left|\frac{\omega_g(k_x)}{k_x}\right| \geq |v_{ph,co}|. \tag{22}$$

where $v_{ph,co} = \omega_g(k_x)/k_x$ is the wave phase velocity in the frame co-moving with the medium. Thus, an electromagnetic instability requires that the relative velocity must exceed the wave velocity of the guided mode. A metal slab supports waves with arbitrarily small wave velocities near the SPP resonance due to a singularity in the density of states. For other types of waveguides the condition $|v| \geq |v_{ph,co}|$ may determine a velocity threshold. For example, for a standard dielectric waveguide with refractive index $n$ one has $|v_{ph,co}| > c/n$ and hence the instabilities can occur only when $|v| \geq c/n$. Interestingly, this threshold is the same as in the Cherenkov problem, which further confirms that the two phenomena are intrinsically related. Furthermore, similar to the Cherenkov problem, the instability threshold is determined by the phase velocity rather than by the group velocity.

## V. Optically induced force

Evidently, to support growing oscillations the system must be pumped somehow. Because there is no explicit optical pump the system is mechanically pumped. Indeed, in the previous calculations the relative velocity of the electric dipole and silver slab was assumed constant. As proven next, this is only possible if an external mechanical force that counteracts the optically induced forces is applied to the pertinent bodies. The optical force acting on the electric dipole is

$$\mathbf{F}_L = \nabla_{\mathbf{r}_0}\left[\mathbf{p}_e \cdot \mathbf{E}\right] + \frac{d}{dt}(\mathbf{p}_e \times \mathbf{B}) \tag{23}$$

where $\mathbf{r}_0$ represents the coordinates of the dipole center of mass [33, 34]. Supposing that the electric dipole oscillates along the *z*-direction, the *x*-component of the force can be decomposed into two terms $F_x = F_{x,1} + F_{x,2}$ where $F_{x,1} = p_e \frac{\partial E_{loc,z}}{\partial x}$, $F_{x,2} = \frac{d}{dt}\left(-p_e B_{loc,y}\right)$, and $E_{loc,z} = E_z^s$ and $B_{loc,y} = B_y^s$ are the local fields (scattered by the moving silver slab) at the position of the dipole. With the help of Eqs. (3)-(4) it is straightforward to evaluate the time averaged $F_{x,1}$ and $F_{x,2}$ for a natural mode of oscillation characterized by the complex-valued frequency (10). The calculations are given in Appendix C. Note that for real-valued frequencies the time-averaged force $F_{x,2}$ vanishes, but for complex-valued frequencies it can be different from zero.

In Fig. 3a we depict the time-averaged force as a function of time for the case of an electric dipole standing at a distance of $1nm$ from a thick silver slab that moves with speed $v = 0.07c$. The time averaging is done over one period of oscillation $T_0 = c/\lambda_0$. As seen, both $F_{x,av,1}$ and $F_{x,av,2}$ are positive and this proves that the optical force acts to drag the dipole towards the direction of motion of the silver slab. In this example, the gradient component of the force ($F_{x,1}$) is several orders of magnitude larger than the component ($F_{x,2}$) associated with the time derivative of the pseudo-momentum [35].

In Fig. 3b and 3c we represent the local field (scattered by the moving silver slab) acting on a dipole driven at $\omega = \omega_{sp}$. As seen, the local field consists of an SPP type wave dragged by the moving slab, such that for $x < 0$ the field is near zero. Clearly, to keep the relative velocity constant one must apply an external force $-F_x$ to the electric dipole and an approximately symmetric force $+F_x$ to the moving silver slab. The work done by the external forces is thus $F_x v > 0$. Evidently, it is this mechanical work that drives the exponentially growing oscillations of the optical field and permits having light generation. In the absence of the external forces (as in a realistic experiment), an optical friction force acts against the relative motion and thus causes a decrease of the relative velocity $v$. In such a case, the field instabilities and the exponential growth are sustained only in the time window wherein $v$ is above the threshold required to have an over-compensation of the radiation loss.

To estimate the strength of the involved forces, the electric dipole moment $p_e$ of the neutral particle at the initial time $t = 0$ is taken equal to the transition dipole moment $d_e$ of the Li I atom [31, 32]. Moreover, the mass of the neutral particle is taken equal to $M \approx 6$ amu, consistent with the atomic mass of Li I. In this case, in the conditions of Fig. 3a the optical force acting on the particle at $t = 0$ is $F_{x,av} = 0.017\,pN$ whereas the optical power pumped into the system is of the order of $F_{x,av} v = 0.35\,\mu W$. The time required to slow down the velocity of the particle to 99% of its initial value is $t_{0.01} \approx 0.01 v/a$ with $a = F_{x,av}/M$ (for simplicity, in this rough estimation $F_{x,av}$ is assumed time independent). This yields $t_{0.01} \approx 5.6 \times 10^7 T_0$, where $T_0$ is the period of oscillation of the emitted radiation. This time window is thus rather large when compared to the period of oscillation.

The force acting on the dipole also has a *z*-component given by $F_z = p_e \frac{\partial E_{loc,z}}{\partial z}$, with a magnitude that is roughly 15% of $F_x$. As shown in Fig. 3a, this force acts to pull the neutral particle towards the silver slab, and hence it is an attractive force analogous to the van der Waals-type forces arising from quantum and thermal fluctuations. The *z*-component of the force also exhibits an exponential growth. From a classical point of view, the emergence of this force is also rather surprising and is another signature of the optical instability. In a realistic experiment, this attractive force may limit the time that a small particle can travel above the metal surface without colliding with it.

## VI. Which of the bodies emits the electromagnetic energy?

As previously discussed, the radiation stress necessarily does some work when the metal slab is sheared with respect to the dipole, and in a closed system this implies a transfer of kinetic energy to the radiation field. Very importantly, the source of the light generation is perceived differently by observers in different reference frames. If the initial electromagnetic field energy is vanishingly small in the reference frame co-moving with a given body (body A) the source of radiation must be the other body (body B). Indeed, only body B has kinetic energy to give away in the frame co-moving with body A. Thus, from a classical point of view, the dynamics of the process is perceived by an observer co-moving with a given object as radiation by the other object in relative motion. This is consistent with the fact that the optical instability results from the interaction of two oscillators (in case of Fig. 1a, the dipole resonator and a surface plasmon) that should be treated on the same footing, and which generate radiation only when they interact with each other.

In particular, from the point of view of the dipole the energy that pumps the system is radiated by the moving slab. A clue for where the energy comes from is given by Eq. (19), which gives the energy density flux ($S_{av}$) created by an incident plane wave with transverse wave vector $(k_x, k_y)$. For an evanescent wave with real-valued frequency $\omega$ impinging on a moving slab (with velocity $v$ in the considered reference frame) $S_{av}$ reduces to:

$$S_{av} = \frac{|E^{inc}|^2}{\eta_0} \frac{1}{\gamma_0} \frac{-\omega}{c} \text{Im}\left\{ R_{p,co}(\omega - vk_x, k_x, k_y) \right\}. \tag{24}$$

The condition (18) ensures that $S_{av} > 0$ when $v = 0$. Surprisingly, one sees that when the velocity is sufficiently large so that $\tilde{\omega} = \omega - vk_x$ and $\omega$ satisfy $\tilde{\omega}\omega < 0$ then $S_{av} < 0$! This striking result confirms that rather than absorbing energy the material medium can generate electromagnetic energy that flows away from the interface towards the source, and explains why it can be seen by the dipole as some kind of energy reservoir. Indeed, when $S_{av} < 0$ the electric dipole can extract energy from the moving medium, and if the distance between the dipole and the moving medium is sufficiently small and the loss is sufficiently low it is possible to have a positive feedback and the generation of growing oscillations. Crucially, this requires that the relative velocity is comparable to the wave phase velocity so that the neutral particle can be coherently pumped by the moving medium. We numerically confirmed that when relativistic corrections are considered it is still possible to have $S_{av} < 0$ [not shown].

Importantly, the possibility of having $S_{av} < 0$ does not imply any intrinsic instability of the moving slab, or that it will start spontaneously emitting light by itself. The instability is only triggered through the interaction with sources (e.g. the electric dipole) that can excite efficiently the evanescent waves that permit $S_{av} < 0$. For example, if in the frame co-moving with a given source there is the opportunity to excite an evanescent wave that creates a negative power flow $S_{av} < 0$ then the instability can be revealed. The condition to have $S_{av} < 0$ is $|v| > |\omega/k_x| = |v_{ph}|$. As previously discussed, for SPPs the phase velocity can be arbitrarily small in the lossless limit, and hence in theory the required velocity to have $S_{av} < 0$ can be much smaller than $c$.

Evidently, if the source is at rest in the frame co-moving with the medium, the condition (18) guarantees that $S_{av} > 0$ for any incident wave. Yet, because an incident evanescent plane wave is also seen in another reference frame as an incident evanescent wave, it may seem paradoxical that $S_{av}$ can switch sign from one reference frame to another.

To make sense of this, let us consider a specific incident evanescent wave such $S_{av} < 0$ in a reference frame where the slab has velocity $v \neq 0$ (for definiteness, we designate this frame as the laboratory frame). Next, we note (similar to what was already discussed in Sect. V) that the incident field can create an $x$-directed force that acts to change the velocity of the material slab. Clearly, in the reference frame co-moving with the slab the energy coming from the source can either be absorbed by the material medium or alternatively it can be used to increase the kinetic energy of the slab. Thus, $S_{av}$ is necessarily non-negative in the co-moving frame. However, in the laboratory frame where $v \neq 0$ the $x$-component of the optical force can do some work, and consequently in a closed system the slab can give away its kinetic energy in the form of radiation leading to $S_{av} < 0$. When the velocity of the slab is enforced to remain constant through the application of an external force, the flux of energy towards the source is still caused by the work done by the optical force.

The optical force can be determined with the Maxwell stress tensor $\overline{\overline{\mathbf{T}}}$ [27]. For a p-polarized incident wave with $k_y = 0$ the nonzero components of the electromagnetic field in the region $z > 0$ (above the metal slab) are $E_x, E_z, H_y$. Thus, in a steady-state ($\omega$ is real-valued), the time-averaged x-component of the Lorentz force per unit of area $F_{x,av}^L / A_0$ acting on the metal semi-space is:

$$\frac{F_{x,av}^L}{A_0} = \frac{\varepsilon_0}{2} \operatorname{Re}\{E_x E_z^*\}. \tag{25}$$

Using $E_x = (1 + R_{p,co}) E^{inc}$ and $E_z = \frac{-ik_x}{\gamma_0}(1 - R_{p,co}) E^{inc}$, it is found that for an incident evanescent wave ($\gamma_0 > 0$):

$$\frac{F_{x,av}^L}{A_0} = \frac{1}{c} \frac{|E^{inc}|^2}{\eta_0} \frac{-k_x}{\gamma_0} \operatorname{Im}\{R_{p,co}(\tilde{\omega}, k_x)\}. \tag{26}$$

The work done by the optical force is $F_{x,av}^L v$ and hence $P_s / A_0 = -v F_{x,av}^L / A_0$ is the power emitted per unit of area due to the conversion of kinetic energy into electromagnetic radiation:

$$\frac{P_s}{A_0} = \frac{1}{c} \frac{|E^{inc}|^2}{\eta_0} \frac{k_x v}{\gamma_0} \operatorname{Im}\{R_{p,co}(\tilde{\omega}, k_x)\}. \tag{27}$$

Note that $P_s = 0$ in the co-moving frame, and hence $P_s$ is observer dependent. The conservation of energy requires that $P_s / A_0 - S_z \geq 0$ where $S_z = -S_{av}$ is the z-component of the Poynting vector at the interface. From Eqs. (18) and (24) it is seen that this condition is indeed satisfied:

$$\frac{P_s}{A_0} - S_z = \frac{1}{c} \frac{|E^{inc}|^2}{\eta_0} \frac{-\tilde{\omega}}{\gamma_0} \text{Im}\{R_{p,co}(\tilde{\omega}, k_x)\} \geq 0. \tag{28}$$

Thus, the finding that the sign of $S_{av}$ can be observer dependent is totally acceptable from a physical point of view, and does not violate energy conservation in any manner. Moreover, this property is absolutely essential so that the interpretation of the phenomenon can be observer dependent, and that an instability is perceived by an observer co-moving with one the bodies as radiation originated in the other body. This is actually not any different from Cherenkov radiation: If an electric charge moves at a certain distance from a metal half-space (similar to Fig. 1a but for a charged particle) the generated light is perceived as being radiated by the charge in the medium co-moving frame. However, for another observer co-moving with the charge the generated light will be perceived as being emitted by the dipoles induced in the moving medium by the static electric field distribution created by the charge. Thus, the direction of energy flow *must* be observer dependent.

## VII. Collective response of many electric dipoles

By considering many identical electric dipoles it may be possible to significantly enhance the Cherenkov-type instabilities, because the optical field emitted by a generic dipole may serve to drive the oscillations of other dipoles. To prove this we consider the scenario of Fig. 1b wherein two identical silver slabs are separated by a distance *d*. Thus, now we have a huge collection of coupled electric dipoles (the silver slab at rest), rather than a single electric dipole. In the non-relativistic regime the natural oscillations of this system can be found by solving the characteristic equation (see Appendix D)

$$1 - e^{-2\gamma_0 d} R_{p,co}(\omega - k_x v) R_{p,co}(\omega) = 0. \tag{29}$$

being $R_{p,co}$ the reflection coefficient introduced previously. Using again the near-field approximation $\gamma_m \approx \gamma_0 \approx k_\parallel$ and Eq. (12), the characteristic equation can be rewritten as

$$1 - e^{-2k_\parallel d} \frac{\omega_{sp}^2}{(\omega - k_x v)(\omega - k_x v + i\Gamma) - \omega_{sp}^2} \frac{\omega_{sp}^2}{\omega(\omega + i\Gamma) - \omega_{sp}^2} = 0, \tag{30}$$

which can be further reduced to a polynomial equation of degree four in $\omega$. We numerically solved this equation with $k_y = 0$ to obtain $\omega = \omega(k_x)$. We found out that for sufficiently large velocities it is possible to have complex-valued solutions $\omega = \omega' + i\omega''$ such that $\omega' \sim \omega_{sp}$ and $\omega''$ can be either positive or negative. The dispersion of the relevant solutions is depicted in Fig. 4a and Fig. 4b [solid lines] for the case wherein $d = 10nm$, $v = 2\omega_{sp}d$ and $\Gamma = 0.2\omega_{sp}$. As seen, the system supports instabilities associated with $\omega'' > 0$, and in addition it supports natural oscillations with finite lifetime ($\omega'' < 0$). This truly remarkable result is robust to changes in the relative velocity and occurs even in case of strong material absorption [Figs. 4c and 4d]. Moreover, relativistic corrections and the effects of time retardation (dashed lines in Fig. 4) result only in small shifts in the dispersion diagrams. The details of the relativistic calculation are given in Appendix D [see Eq. (D1)].

## VIII. Conclusion

In summary, we theoretically demonstrated that two neutral closely spaced polarizable bodies in relative motion can start spontaneously emitting light if their relative velocity is sufficiently large, even if the initial optical field is vanishingly small. This effect occurs due to the conversion of kinetic energy into optical energy, similar to the well known Vasilov-Cherenkov effect, but here for neutral matter. This instability can be understood as being the result of the hybridization of resonances with frequencies (as seen in the respective co-moving frames) with opposite signs. While in the Cherenkov effect the recoil force is constant, in our system, quite dramatically, the force builds up exponentially with time. The exponentially growing oscillations may lead to the emergence of strong nonlinear effects, and for sufficiently large field amplitudes to tunnel ionization. Moreover, as illustrated by the example of Fig. 4, the reported instabilities occur even in case of a "continuous beam" of moving particles, quite different from the Cherenkov phenomenon which takes place only for modulated beams. Crucially, the analysis of this article is *completely classical*, but the reported effects may lead to exciting developments in the framework of quantum electrodynamics, particularly in context of noncontact quantum friction [23-26]. Furthermore, the present theory (see also Ref. [24]) raises the question if there is anything specifically "quantum" in some phenomena involving the quantum vacuum [10]-[23], and suggests that "quantum friction" and related effects may eventually be explained with classical arguments with the additional ingredient of a spectrum of random electromagnetic radiation [36].

**Acknowledgement:** This work is supported in part by Fundação para a Ciência e a Tecnologia grant number PTDC/EEI-TEL/2764/2012.

## *Appendix A: The reflection matrix*

Here, we derive the reflection matrix $\mathbf{R}(\omega, k_x, k_y)$ for a moving slab surrounded by a vacuum. The reflection matrix is such that for an incident plane wave propagating in the vacuum region with transverse wave vector $(k_x, k_y)$ and transverse fields $E_x^{inc}, E_y^{inc}$, the corresponding reflected fields satisfy:

$$\begin{pmatrix} E_x^{ref} \\ E_y^{ref} \end{pmatrix} = \mathbf{R}(\omega, k_x, k_y) \cdot \begin{pmatrix} E_x^{inc} \\ E_y^{inc} \end{pmatrix}. \tag{A1}$$

It is assumed that the interface is normal to the *z*-direction so that the transverse components of the electric field are tangential to the interface.

Suppose that the pertinent body moves with velocity $\mathbf{v} = v\hat{\mathbf{x}}$ in the laboratory frame. It is possible to relate $\mathbf{R}(\omega, k_x, k_y)$ with $\mathbf{R}_{co}(\tilde{\omega}, \tilde{k}_x, \tilde{k}_y)$, which is defined in the same manner as $\mathbf{R}(\omega, k_x, k_y)$ but for the frame co-moving with the body:

$$\begin{pmatrix} \tilde{E}_x^{ref} \\ \tilde{E}_y^{ref} \end{pmatrix} = \mathbf{R}_{co}(\tilde{\omega}, \tilde{k}_x, \tilde{k}_y) \cdot \begin{pmatrix} \tilde{E}_x^{inc} \\ \tilde{E}_y^{inc} \end{pmatrix}. \tag{A2}$$

All the quantities with the tilde hat are calculated in the co-moving frame. The electromagnetic fields in the two frames are related by [27]:

$$\tilde{E}_x = E_x \tag{A3a}$$

$$\tilde{E}_y = g(E_y - vB_z) \tag{A3b}$$

where $g = 1/\sqrt{1-\beta^2}$ with $\beta = v/c$. In the vacuum region $i\omega B_z = \frac{\partial E_y}{\partial x} - \frac{\partial E_x}{\partial y}$ and hence for a plane wave, $B_z = -\frac{k_y}{\omega} E_x + \frac{k_x}{\omega} E_y$. Thus, it follows that:

$$\begin{pmatrix} \tilde{E}_x \\ \tilde{E}_y \end{pmatrix} = \begin{pmatrix} E_x \\ E_y \end{pmatrix} + g \underbrace{\begin{pmatrix} 0 & 0 \\ \beta c \frac{k_y}{\omega} & 1 - \frac{1}{g} - \beta c \frac{k_x}{\omega} \end{pmatrix}}_{\mathbf{A}(\omega, k_x, k_y, v)} \begin{pmatrix} E_x \\ E_y \end{pmatrix} \tag{A4}$$

Substituting this result into Eq. (A2), it is found that:

$$(1+\mathbf{A}) \cdot \begin{pmatrix} E_x^{ref} \\ E_y^{ref} \end{pmatrix} = \mathbf{R}_{co,i}\left(\tilde{\omega}, \tilde{k}_x, \tilde{k}_y\right) \cdot (1+\mathbf{A}) \cdot \begin{pmatrix} E_x^{inc} \\ E_y^{inc} \end{pmatrix} \qquad (A5)$$

where $\left(\tilde{\omega}, \tilde{k}_x, \tilde{k}_y\right)$ are related to the corresponding parameters in the laboratory frame through the relativistic Doppler shift formulas [27]:

$$\tilde{\omega} = g(\omega - vk_x), \qquad \tilde{k}_x = g(k_x - \omega v/c^2), \qquad \tilde{k}_y = k_y. \qquad (A6)$$

Comparing Eqs. (A1) and (A5) we find that:

$$\mathbf{R}(\omega, k_x, k_y) = (1+\mathbf{A})^{-1} \cdot \mathbf{R}_{co}\left(\tilde{\omega}, \tilde{k}_x, \tilde{k}_y\right) \cdot (1+\mathbf{A}) \qquad (A7)$$

For the case of an unbounded semi-infinite metal slab, it can be shown using vector transmission line theory that the matrix $\mathbf{R}_{co}$ is given by:

$$\mathbf{R}_{co}(\omega, k_x, k_y) = \left(\mathbf{Y}_{c0}^+ + \mathbf{Y}_{c,d}^+\right)^{-1} \cdot \left(\mathbf{Y}_{c0}^+ - \mathbf{Y}_{c,d}^+\right), \qquad (A8)$$

where $\mathbf{Y}_c^+$ is defined by

$$\mathbf{Y}_c^+ = \frac{c}{\omega\mu} \frac{1}{i\gamma} \begin{pmatrix} k_x^2 - \gamma^2 & k_y k_x \\ k_y k_x & k_y^2 - \gamma^2 \end{pmatrix}, \qquad \gamma^2 = k_x^2 + k_y^2 - \varepsilon\mu\omega^2/c^2 \qquad (A9)$$

where $\varepsilon$ and $\mu$ are the relative permittivity and permeability, respectively, and $\gamma$ is the propagating constant for the +z direction (this determines the branch cut in the definition of $\gamma$). The subscripts "$0$" and "$d$" indicate if $\varepsilon$ and $\mu$ are either calculated in the vacuum region ($\varepsilon = \mu = 1$) or in metal region ($\varepsilon = \varepsilon_m$ and $\mu = \mu_m$). Formula (A8) is valid independent of the direction of propagation of the incoming wave. Thus, using Eqs. (A7) and (A8) one can easily determine the reflection matrix $\mathbf{R}$ in the laboratory frame. In the non-relativistic limit $\mathbf{A} \approx 0$ in Eq. (A7), and hence we obtain:

$$\mathbf{R}(\omega, k_x, k_y) \approx \mathbf{R}_{co}(\omega - vk_x, k_x, k_y), \qquad v/c \ll 1. \qquad (A10)$$

The eigenvalues $R_{p,co}$ and $R_{s,co}$ of $\mathbf{R}_{co}$ are given by:

$$R_{l,co}(\omega, k_x, k_y) = \frac{Y_0^l - Y_d^l}{Y_0^l + Y_d^l}, \qquad l=s,p. \tag{A11}$$

where $Y^p = \frac{\varepsilon}{i\gamma}\frac{\omega}{c}$ and $Y^s = \frac{c}{\omega\mu}i\gamma$ are the normalized wave admittances for $p$ and $s$ polarized waves. The corresponding eigenvectors are such that the incoming wave has electric field either parallel or perpendicular to the plane of incidence: $p$ and $s$ type polarizations.

## *Appendix B: Natural oscillations of a dipole with an isotropic response*

In the main text, it is assumed that the dipole response is anisotropic so that it can only oscillate along the $z$-direction. Next, we explain how the theory is modified in case of a dipole with an isotropic response and estimate the corresponding threshold for an optical instability.

In the general the local electric field acting on the dipole is related to the electric dipole moment $\mathbf{p}_e$ as $\mathbf{E}_{loc} = \mathbf{C}_{int} \cdot \mathbf{p}_e / \varepsilon_0$ where the interaction constant $\mathbf{C}_{int}$ is now a 3×3 tensor. Proceeding as in the main text, it can be shown that:

$$\mathbf{C}_{int} = \frac{1}{(2\pi)^2} \iint dk_x dk_y \left[\mathbf{1}_t + \hat{\mathbf{z}}\frac{i\mathbf{k}_t}{\gamma_0}\right] \cdot \mathbf{R}(\omega, k_x, k_y) \cdot \left[i\mathbf{k}_t\gamma_0\hat{\mathbf{z}} + \left(\frac{\omega^2}{c^2}\right)\mathbf{1}_t - \mathbf{k}_t\mathbf{k}_t\right] \frac{e^{-2\gamma_0 d}}{2\gamma_0}, \tag{B1}$$

where $\mathbf{1}_t = \hat{\mathbf{x}}\hat{\mathbf{x}} + \hat{\mathbf{y}}\hat{\mathbf{y}}$ is the transverse identity matrix and $\mathbf{uv} \equiv \mathbf{u} \otimes \mathbf{v}$ represents the tensor product of two generic vectors $\mathbf{u}$ and $\mathbf{v}$. Note that $\hat{\mathbf{z}} \cdot \mathbf{C}_{int} \cdot \hat{\mathbf{z}}$ is the interaction constant (6) calculated in Sect. II. The natural oscillations of the dipole are now determined by the homogeneous system:

$$\left[\alpha_e^{-1}(\omega) - \mathbf{C}_{int}(\omega)\right] \cdot \mathbf{p}_e = 0. \tag{B2}$$

which leads to the dispersion equation $\det(\alpha_e^{-1}(\omega) - \mathbf{C}_{int}(\omega)) = 0$. For near field interactions it can be assumed that $k_\parallel = |\mathbf{k}_t| \gg \omega/c$, so that $\gamma_0 \approx k_\parallel$. This gives:

$$\mathbf{C}_{int} \approx \frac{1}{(2\pi)^2} \iint dk_x dk_y \left[\mathbf{1}_t + i\hat{\mathbf{z}}\hat{\mathbf{k}}_t\right] \cdot \mathbf{R}(\omega, k_x, k_y) \cdot \left[i\hat{\mathbf{k}}_t \hat{\mathbf{z}} - \hat{\mathbf{k}}_t \hat{\mathbf{k}}_t\right] \frac{k_\parallel}{2} e^{-2k_\parallel d}, \quad (B3)$$

where $\hat{\mathbf{k}}_t = \mathbf{k}_t / |\mathbf{k}_t|$. Furthermore, in the non-relativistic regime one can use the approximation (A10). Taking into account that the eigenvalues of $\mathbf{R}_{co}$ are the reflection coefficients for $p$ and $s$ polarized waves $R_{p,co}$ and $R_{s,co}$ we see that:

$$\mathbf{R}_{co} = R_{p,co} \hat{\mathbf{k}}_t \otimes \hat{\mathbf{k}}_t + R_{s,co} (\hat{\mathbf{z}} \times \hat{\mathbf{k}}_t) \otimes (\hat{\mathbf{z}} \times \hat{\mathbf{k}}_t) \quad (B4)$$

Hence, in the non-relativistic regime we obtain the following simplified formula for the interaction tensor:

$$\mathbf{C}_{int}(\omega) \approx \frac{-1}{(2\pi)^2} \iint dk_x dk_y R_{p,co}(\tilde{\omega}, k_x, k_y) \frac{k_\parallel}{2} e^{-2k_\parallel d} \left[\hat{\mathbf{k}}_t \otimes \hat{\mathbf{k}}_t + \hat{\mathbf{z}} \otimes \hat{\mathbf{z}} - i\hat{\mathbf{k}}_t \otimes \hat{\mathbf{z}} + i\hat{\mathbf{z}} \otimes \hat{\mathbf{k}}_t\right], \quad (B5)$$

where $\tilde{\omega} = \omega - vk_x$. Because $R_{p,co}(\tilde{\omega}, k_x, k_y)$ is an even function of $k_y$ it is clear that $C_{int,xy} = C_{int,yx} = C_{int,yz} = C_{int,zy} = 0$, where $C_{int,ij}$ represents the $ij$ element of the tensor. Therefore, for oscillations with $\mathbf{p}_e = p_{e,x}\hat{\mathbf{x}} + p_{e,z}\hat{\mathbf{z}}$ the homogeneous equation (B2) reduces to:

$$\begin{pmatrix} C_{int,xx} - \alpha_e^{-1} & C_{int,xz} \\ C_{int,zx} & C_{int,zz} - \alpha_e^{-1} \end{pmatrix} \begin{pmatrix} p_{e,x} \\ p_{e,z} \end{pmatrix} = 0. \quad (B6)$$

As discussed in Sect. IV, $\mathbf{C}_{int}(\omega)$ is mainly determined by the poles of the reflection coefficient, specifically by the guided modes with $k_y \approx 0$. The integrand of (B5) when evaluated in the vicinity of a pole with $(k_x, k_y) \approx (k_{x,g}, 0)$ has the following symmetries:

$c_{\text{int},xx} \approx c_{\text{int},zz}$ and $c_{\text{int},xz} = -c_{\text{int},zx} \approx -is\, c_{\text{int},zz}$. Here, $c_{\text{int},ij}$ denotes the $ij$ element of the integrand of (B5) and $s = \text{sgn}(k_{x,g})$. Based on this discussion, we can obtain the following estimations for the interaction constant elements $C_{\text{int},xx} \approx C_{\text{int},zz} = C_{\text{int}}$ and $C_{\text{int},xz} = -C_{\text{int},zx} \approx -is\, C_{\text{int}}$, where $C_{\text{int}}$ is the interaction constant calculated in the main text [Eq. (6)] and $s = \text{sgn}(k_{x,g})$ is determined by the dominant pole. Within the validity of these approximations, nontrivial solutions of Eq. (B6) can occur only when $\left(C_{\text{int}} - \alpha_e^{-1}\right)^2 - C_{\text{int}}^2 = 0$, which is equivalent to say that either $2C_{\text{int}} - \alpha_e^{-1} = 0$ or $\alpha_e^{-1} = 0$. The first case is the interesting one and corresponds to oscillations with $\mathbf{p}_e = p_e\left(\hat{\mathbf{x}} + is\hat{\mathbf{z}}\right)$, i.e. to a circular polarization. This is consistent with the fact that highly confined SPPs are circularly polarized. Thus, within the validity of our analysis, in the isotropic case the interaction constant is multiplied by a factor of two, as compared to the anisotropic case. In particular, for $\omega = \omega_{sp}$, $v = 5.26\omega_{sp}d$, and $\Gamma = 0^+$ the condition to have an optical instability is now $-1/6\pi\left(\omega_{sp}/c\right)^3 + 2\times10^{-4}/d^3 > 0$, which for a silver film imposes that $d < 11.2nm$, which is a slightly less restrictive than in the anisotropic case.

## *Appendix C: The optical force on the electric dipole*

As discussed in the main text, the instantaneous optical force acting on the electric dipole satisfies:

$$F_x = \frac{d}{dt}\left(-p_e B_{loc,y}\right) + p_e \frac{\partial E_{loc,z}}{\partial x} \tag{C1}$$

where $E_{loc,z} = E_z^s$ and $B_{loc,y} = B_y^s$. For time harmonic fields (with time variation of the type $e^{-i\omega t}$) the fields scattered by the moving slab are given by Eqs. (3)-(4). Hence, using $\nabla \times \mathbf{E} = i\omega\mathbf{B}$ we obtain $cB_y = \frac{1}{ik_0}(\partial_z E_x - \partial_x E_z)$ with $k_0 = \omega/c$, so that the scattered magnetic field is:

$$cB_y^s = \frac{1}{k_0}\frac{p_e}{\varepsilon_0}\frac{1}{(2\pi)^2}\iint dk_x dk_y \left[-\gamma_0^2 \hat{\mathbf{x}} \cdot \mathbf{R}(\omega,k_x,k_y) \cdot \mathbf{k}_t + k_x \mathbf{k}_t \cdot \mathbf{R}(\omega,k_x,k_y) \cdot \mathbf{k}_t\right]\frac{e^{-\gamma_0(d+z)}}{2\gamma_0}e^{i(k_x x + k_y y)} \quad (C2)$$

Let us introduce constants $C_{B,\text{int}}$ and $\tilde{C}_{\text{int},x}$ such that the fields calculated at the position of the dipole satisfy $\frac{\partial E_{loc,z}}{\partial x} = \tilde{C}_{\text{int},x}\frac{p_e}{\varepsilon_0}$ and $cB_{loc,y} = C_{B,\text{int}}\frac{p_e}{\varepsilon_0}$. It is evident that:

$$\tilde{C}_{\text{int},x} = -\frac{1}{(2\pi)^2}\iint dk_x dk_y \, \mathbf{k}_t \cdot \mathbf{R}(\omega,k_x,k_y) \cdot \mathbf{k}_t \frac{ik_x}{2\gamma_0} e^{-2\gamma_0 d} \quad (C3)$$

$$C_{B,\text{int}} = \frac{1}{k_0}\frac{1}{(2\pi)^2}\iint dk_x dk_y \left[-\gamma_0^2 \hat{\mathbf{x}} \cdot \mathbf{R}(\omega,k_x,k_y) \cdot \mathbf{k}_t + k_x \mathbf{k}_t \cdot \mathbf{R}(\omega,k_x,k_y) \cdot \mathbf{k}_t\right]\frac{1}{2\gamma_0}e^{-2\gamma_0 d}. \quad (C4)$$

In a non-relativistic approximation [Eq. (A10)], it is possible to write:

$$\tilde{C}_{\text{int},x} \approx -\frac{1}{(2\pi)^2}\iint dk_x dk_y \frac{ik_x k_\parallel^2}{2\gamma_0} R_{p,co}(\omega - vk_x, k_x, k_y) e^{-2\gamma_0 d}, \quad (C5)$$

$$C_{B,\text{int}} \approx \frac{1}{(2\pi)^2}\iint dk_x dk_y \frac{k_0 k_x}{2\gamma_0} R_{p,co}(\omega - vk_x, k_x, k_y) e^{-2\gamma_0 d}. \quad (C6)$$

Thus, the instantaneous optical force acting on the dipole for a real-valued time harmonic field associated with the complex-valued oscillation frequency $\omega = \omega' + i\omega''$ is:

$$\begin{aligned}F_x &= \text{Re}\{p_e e^{-i\omega t}\}\text{Re}\left\{\tilde{C}_{\text{int},x}\frac{p_e}{\varepsilon_0}e^{-i\omega t}\right\} - \frac{1}{c}\frac{d}{dt}\left(\text{Re}\{p_e e^{-i\omega t}\}\text{Re}\left\{C_{B,\text{int}}\frac{p_e}{\varepsilon_0}e^{-i\omega t}\right\}\right)\\ &= \frac{p_e^2}{\varepsilon_0 d^4}\left[e^{2\omega'' t}\text{Re}\{e^{-i\omega' t}\}\text{Re}\{\tilde{C}_{\text{int},x}d^4 e^{-i\omega' t}\} - \frac{d}{c}\frac{d}{dt}\left(e^{2\omega'' t}\text{Re}\{e^{-i\omega' t}\}\text{Re}\{C_{B,\text{int}}d^3 e^{-i\omega' t}\}\right)\right]\end{aligned}$$
(C7)

where we assumed without loss of generality that $p_e$ is real-valued. The force $F_x$ changes very rapidly on the scale of one period $T_0 = 2\pi/\omega'$. For $\omega'' \ll \omega'$ the force averaged force a time scale of the order of $T_0$ is given by:

$$F_{x,av} = \frac{p_e^2}{\varepsilon_0 d^4} e^{2\omega''t} \left( \frac{1}{2} \text{Re}\{\tilde{C}_{int,x} d^4\} - \frac{\omega'' d}{c} \text{Re}\{C_{B,int} d^3\} \right). \tag{C8}$$

Similarly, the $z$-component of the force $F_z = p_e \dfrac{\partial E_{loc,z}}{\partial z}$ can be written as

$$F_z = \text{Re}\{p_e e^{-i\omega t}\} \text{Re}\left\{\tilde{C}_{int,z} \frac{p_e}{\varepsilon_0} e^{-i\omega t}\right\} \text{ with:}$$

$$\tilde{C}_{int,z} = \frac{1}{(2\pi)^2} \iint dk_x dk_y \mathbf{k}_t \cdot \mathbf{R}(\omega, k_x, k_y) \cdot \mathbf{k}_t \frac{1}{2} e^{-2\gamma_0 d}. \tag{C9}$$

Thus, the time-averaged $F_z$ is given by:

$$F_{z,av} = \frac{p_e^2}{\varepsilon_0 d^4} e^{2\omega''t} \frac{1}{2} \text{Re}\{\tilde{C}_{int,z} d^4\}. \tag{C10}$$

## *Appendix D: Natural oscillations in a cavity formed by two bodies in relative motion*

Here, we obtain the dispersion equation for the natural modes of oscillation in a cavity formed by two bodies in relative motion separated by a vacuum gap [Fig. 1b]. Let $\mathbf{R}_i(\omega, k_x, k_y)$ ($i$=1,2) be the reflection matrix associated with the $i$-th body, defined as in Appendix A. Evidently, the natural modes of oscillation satisfy the characteristic equation $D(\omega, k_x, k_y) = 0$ with

$$D(\omega, k_x, k_y, v_1, v_2) = \det(\mathbf{1} - e^{-2\gamma_0 d} \mathbf{R}_1 \cdot \mathbf{R}_2), \tag{D1}$$

where $\gamma_0 = \sqrt{k_x^2 + k_y^2 - \omega^2/c^2}$, **1** is the identity matrix, and $d$ is the distance between the moving bodies. The matrices $\mathbf{R}_1$ and $\mathbf{R}_2$ are computed using Eqs. (A7) and (A8). It can be checked that $D(\omega, k_x, k_y, v_1, v_2)$ is relativistically invariant, i.e. it has the same value in any frame when the frequency, the wave vector, and the velocities are relativistically transformed.

In the non-relativistic limit, $v_i/c \ll 1$, $\mathbf{R}_1$ and $\mathbf{R}_2$ are diagonalizable in the same basis [see Eq. (A10)] and hence the characteristic equation for the natural modes of oscillation [Eq. (D1)] reduces to $D^p(\omega, k_x, k_y) D^s(\omega, k_x, k_y) = 0$ with,

$$D^l(\omega, k_x, k_y) = 1 - e^{-2\gamma_0 d} R_{co,1}^l(\omega - v_1 k_x, k_x, k_y) R_{co,2}^l(\omega - v_2 k_x, k_x, k_y), \quad l=s,p \quad (D2)$$

where $R_{co,1}^l$ represent the eigenvalues of the reflection matrices [Eq. (A11)].

[32] The response of a neutral atom can be modeled semi-classically by $\alpha_e^{-1} \approx \frac{\hbar \omega_0 \varepsilon_0}{d_e^2}\left[\frac{1}{2\omega_0^2}\left(\gamma^2 + \omega_0^2 - \omega^2\right) - i\gamma \frac{\omega}{\omega_0^2}\right]$ [30], with $\omega_0$ the transition frequency, $d_e$ is the real transition dipole moment, $\gamma = R_{sp}/2$, and $R_{sp} = \frac{1}{3\pi\varepsilon_0 \hbar} d_e^2 \left(\frac{\omega_0}{c}\right)^3$ is the spontaneous emission rate (unlike Ref. [30] we do not consider the effect of orientation averaging). Using $Q = \omega_0 / R_{sp}$ it can be checked that this polarizability response is consistent with that considered in the main text for $\omega \sim \omega_0$. The Li I atom has a resonance (corresponding to the lowest energy transition from the ground state) at $\lambda_0 = 671 nm$ and $R_{sp} = 0.37 \times 10^8 s^{-1}$ [31].

[33] J. P. Gordon, *Phys. Rev. A* **8**, 14 (1973).

[34] E. A. Hinds, S. M. Barnett, *Phys. Rev. Lett.* **102**, 050403 (2009).

[35] R. Loudon, L. Allen, D. F. Nelson, *Phys. Rev. E*, **55**, 1071 (1997).

[36] T. H. Boyer, *Am. J. Phys.* **79**, 1163 (2011).

# *Figures*

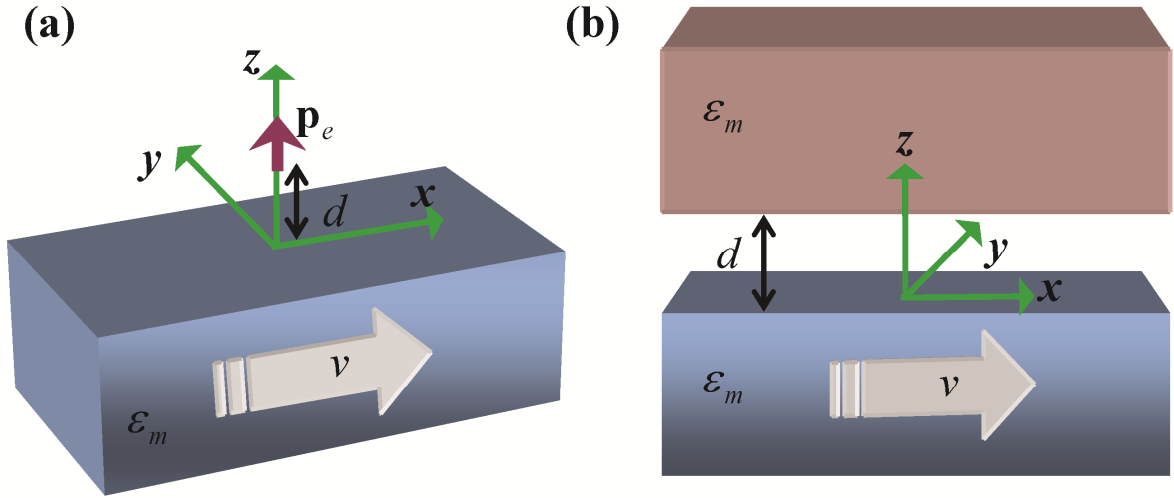

Fig. 1. (Color online) **(a)** A vertical electric dipole stands above a semi-infinite metallic region moving with a relative velocity *v*. **(b)** Two metallic semi-infinite spaces are separated by a vacuum gap with thickness *d*. The lower region moves with a relative velocity *v*.

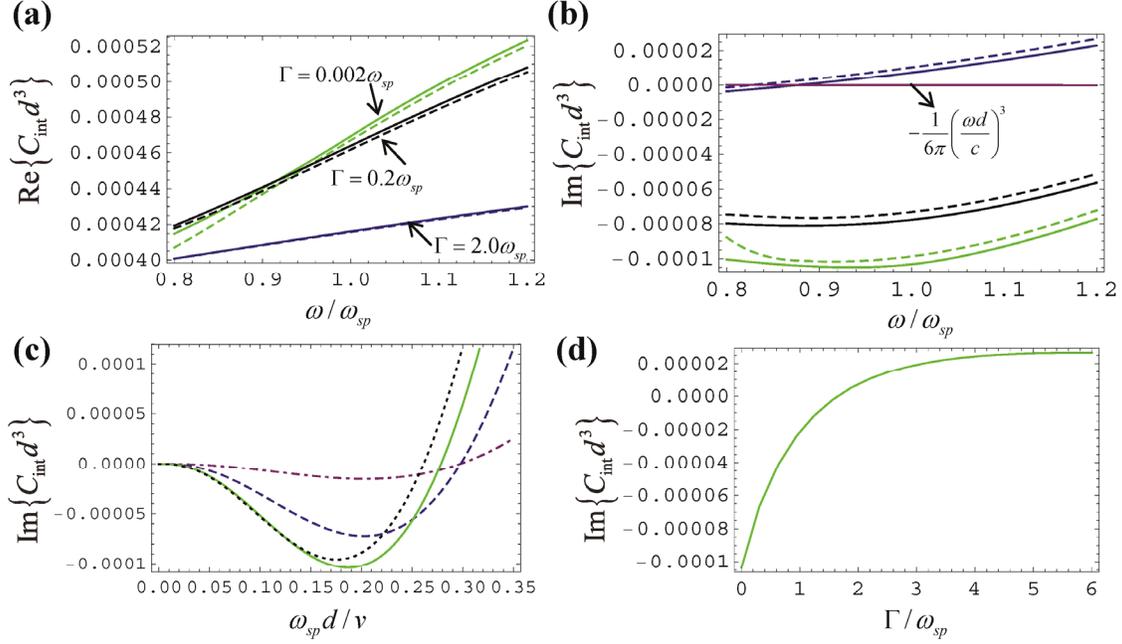

Fig. 2. (Color online) **(a)** Real part and **(b)** imaginary part of the interaction constant as a function of frequency for a point dipole placed at a distance $d = 1nm$ [$\omega_{sp}d/c = 0.0135$] above a silver semi-space in relative motion. The relative velocity of the silver semi-space is $v = 5.26\omega_{sp}d = 0.07c$. Green lines: $\Gamma = 0.002\omega_{sp}$. Black lines: $\Gamma = 0.2\omega_{sp}$. Blue lines: $\Gamma = 2.0\omega_{sp}$. The solid lines represent the non-relativistic calculation [Eq. (13)] and the dashed lines the exact relativistic calculation. **(c)** imaginary part of the interaction constant as a function of $\omega_{sp}d/v$ in the non-relativistic lossless limit ($\Gamma = 0^+$). Purple dot-dashed line: $\omega = 0.1\omega_{sp}$; Blue dashed line: $\omega = 0.5\omega_{sp}$; Green solid line: $\omega = 1.0\omega_{sp}$; Black dotted line: $\omega = 1.1\omega_{sp}$; **(d)** imaginary part of the interaction constant at $\omega = \omega_{sp}$ [for $d = 1nm$ and $v = 5.26\omega_{sp}d$] as a function of the normalized collision frequency $\Gamma$ in the non-relativistic limit [Eq. (13)].

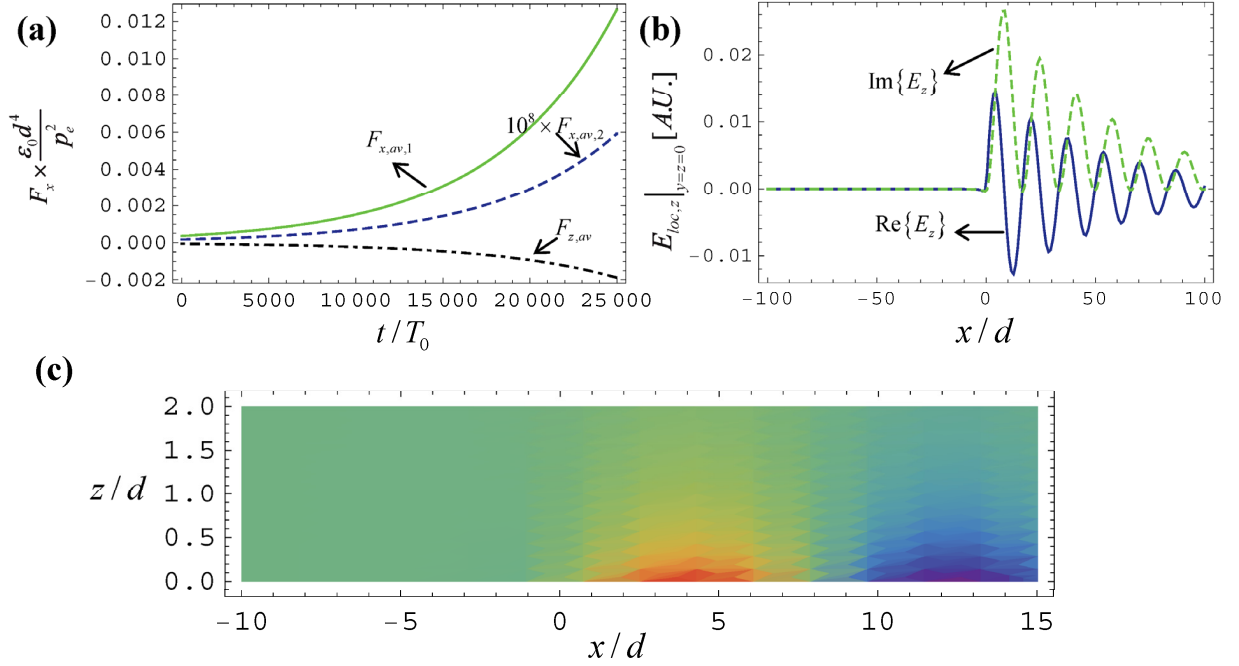

Fig. 3. (Color online) **(a)** Normalized friction force as a function of time acting on an electric dipole standing $1nm$ [$\omega_{sp}d/c = 0.0135$] above a thick silver slab. $F_{x,av,1}$ and $F_{x,av,2}$ represent the two components of the time averaged optical force (see the main text), and $F_{z,av}$ is the time averaged $z$-component of the force. The velocity of the silver slab is $v = 5.26\omega_{sp}d = 0.07c$. The time axis is normalized to $T_0 = \lambda_0/c$ where $\lambda_0 = 671nm$ is the resonant wavelength of the electric dipole. **(b)** $z$-component of the local field (backscattered by the moving silver slab) at $y = z = 0$ for an electric dipole positioned at $(0,0,d)$ and oscillating with $\omega = \omega_{sp}$. **(c)** Density plot of the real part of the local field for the same scenario as in (b). Reddish (bluish) colors represent positive (negative) values of the electric field. Greenish colors (as in the region $x < 0$) represent field amplitudes near zero.

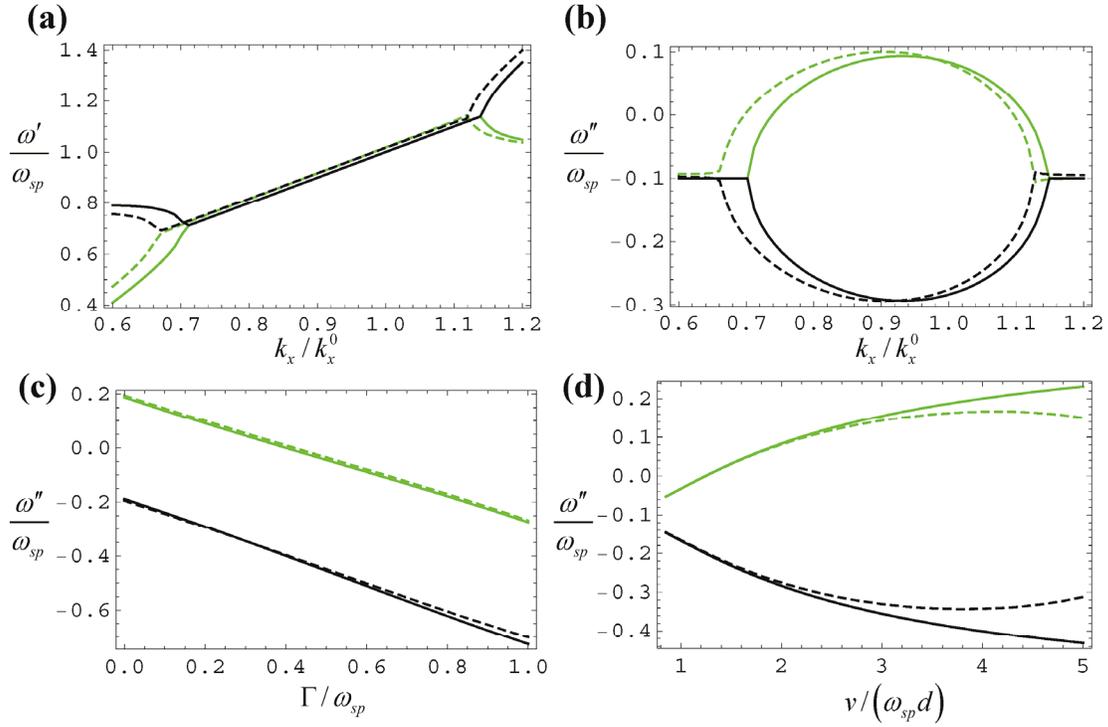

Fig. 4. (Color online) Two silver semi-spaces separated by $d = 10nm$ [$\omega_{sp}d/c = 0.135$] are in relative motion. **(a)** Real part and **(b)** imaginary part of oscillation frequencies of two "twin" natural modes (black and green lines) as a function of the normalized $k_x$ for $\Gamma = 0.2\omega_{sp}$, and $v = 2\omega_{sp}d$. The transverse wave number is normalized to $k_x^0 = 2\omega_{sp}/v$. **(c)** *Effect of loss:* Imaginary part of the natural oscillation frequencies as a function of the normalized collision frequency $\Gamma$ of silver for $k_x = 0.9k_x^0$. **(d)** *Effect of changing the relative velocity:* Imaginary part of the natural oscillation frequencies as a function of the relative velocity for $k_x = 2\omega_{sp}/v$ and $\Gamma = 0.2\omega_{sp}$. In all the plots, $k_y = 0$, and the solid lines represent the non-relativistic calculation and the dashed lines the exact relativistic calculation.